\documentclass[10pt,journal,compsoc]{IEEEtran}

\usepackage[normalem]{ulem}
\usepackage{amsmath}
\usepackage{pifont}
\newcommand{\cmark}{\ding{51}}%
\usepackage{algorithmic}
\usepackage{graphicx}
\usepackage{textcomp}
\usepackage{xcolor}

\usepackage{subcaption}

\usepackage[linesnumbered,ruled,vlined]{algorithm2e}
\usepackage{pgfplots}
\pgfplotsset{compat=1.16}

\usepackage{tabulary}
\usepackage{multirow}

\usepackage{hyperref}

\usepackage{cleveref}

\definecolor{light-gray}{gray}{0.95}

\newcommand{\code}[1]{\colorbox{light-gray}{\texttt{#1}}}
\newcommand{\comp}{\texttt}

\SetCommentSty{mycommfont}

\newboolean{showcomments}
\setboolean{showcomments}{false}
\ifthenelse{\boolean{showcomments}}
{ \newcommand{\mynote}[3]{
    \fbox{\bfseries\sffamily\scriptsize#1}
    {\small$\blacktriangleright$\textsf{\emph{\color{#3}{#2}}}$\blacktriangleleft$}}}
{ \newcommand{\mynote}[3]{}}

\newcommand{\jpb}[1]{\mynote{jpb}{#1}{red}}
\newcommand{\rr}[1]{\mynote{rr}{#1}{orange}}
\newcommand{\jcm}[1]{\mynote{jcm}{#1}{brown}}
\newcommand{\igk}[1]{\mynote{igk}{#1}{magenta}}

\newcommand{\name}{HyPlacer\xspace}
\newcommand{\optane}{DCPMM\xspace}

\title{Dynamic Page Placement on Real Persistent Memory Systems}

\author{\IEEEauthorblockN{Miguel~Marques, Ilia Kuzmin, João Barreto, José Monteiro, and Rodrigo Rodrigues}\\
\IEEEauthorblockA{INESC-ID, Instituto Superior Técnico, Universidade de Lisboa\\
\{miguel.soares.marques, ilia.kuzmin, joao.barreto\}@tecnico.ulisboa.pt, jcm@inesc-id.pt, rodrigo.miragaia.rodrigues@tecnico.ulisboa.pt}}

\begin{document}
\date{}

\thispagestyle{empty}


\maketitle

\begin{abstract}

As persistent memory (PM) technologies emerge, hybrid memory architectures combining DRAM with PM bring the potential to provide a tiered, byte-addressable main memory of unprecedented capacity.
%
Nearly a decade after the first proposals for these hybrid architectures, 
the real technology has finally reached commercial availability with Intel Optane™ DC Persistent Memory (\optane).
This raises the challenge of designing systems that realize this potential in practice, namely through effective approaches that dynamically decide at which memory tier should pages be placed.

In this paper, we are the first, to our knowledge, to systematically
analyze
tiered page placement on real \optane-based systems.
To this end, we start by revisiting the assumptions of state-of-the-art proposals, and confronting them with the idiosyncrasies of today's off-the-shelf \optane-equipped architectures.
This empirical study reveals that some of the key design choices in the literature rely on important assumptions that are not verified in present-day
DRAM-\optane memory architectures.

Based on the lessons from this study, we design and implement \name, a tool for tiered page placement in \emph{off-the-shelf} Linux-based systems equipped with DRAM+\optane.
In contrast to previous proposals, \name follows
an approach guided by
two main practicality \jcm{shouldn't this be just "practical"}
principles:
1) it is tailored to the performance idiosyncrasies of off-the-shelf DRAM+\optane systems;
and 2) it can be seamlessly integrated into Linux with minimal kernel-mode components, while ensuring extensibility to other HMAs and other data placement policies. Our experimental evaluation of \name\ shows that it outperforms both solutions proposed in past literature and placement options that are currently available in off-the-shelf \optane-equipped Linux systems, reaching an improvement of up to 11x when compared to the default memory policy in Linux.



\end{abstract}

\section{Introduction}
Today's data-intensive applications are characterized by a high degree of data complexity and parallelism, as well as an ever-increasing demand for memory capacity. \jpb{it would be nice to have a citation supporting the previous claim. if not, we can simply keep it citation-less.}
However, scaling up DRAM memory to address such trends is rapidly hitting a well-known wall -- the amount of available DRAM is limited by power, cooling, area constraints, and cost \cite{10.1109/ISCA45697.2020.00061}.

In this context, the emerging persistent memory (PM) technologies can become a
game changer, 
not so much for their most highlighted property -- persistence --, but for an unprecedented
combination of other properties:
PM delivers read/write latencies in the nanosecond range;
it is byte-addressable, allowing applications to directly access them by issuing load and store instructions; and,
compared to DRAM, PM is denser, providing a higher capacity per module, while offering lower cost per bit.

As a consequence, a natural trend is to have heterogeneous memory hierarchies (HMHs), which
combine both memory technologies, PM and DRAM. This enables a
\emph{tiered} main memory, where applications can place
and process much larger working sets than in DRAM-only systems, therefore minimizing accesses to local or distributed storage. Such HMHs are useful even for volatile working sets, which are the focus of our paper.



The performance duality in DRAM+PM HMHs raises a challenging data placement problem: \emph{at which tier in the HMH should the data objects of an application be placed for optimal performance?}
Allocating data to the appropriate memory, taking into account the performance differences between both tiers, becomes a decisive challenge to the effective scalability of data-intensive applications.


These data placement strategies received considerable attention from the research community in the last decade, following the initial promises of new PM technologies and their integration into DRAM+PM HMHs.
However, the proposals that resulted from these initial research efforts suffer from the intrinsic limitation of not having access to commercial PM hardware, since it was not available at the time.  In particular, we identify two fundamental shortcomings in this initial research.

First, 
these proposals are designed based on
conjectural
assumptions about the performance of then
future PM-based systems.
Furthermore, their experimental evaluation relies on simulation or emulation methods that
fail to capture performance intricacies that are only noticeable with the real technology \cite{optane_study4}.

Second, with a few exceptions, the early research work generally overlooks the issue of how the proposed techniques are to be integrated into
the system stack of PM-based systems.
As we detail later on, some papers simply propose placement algorithms, regardless of how they can be implemented within the complex system stack of today’s systems; other proposals resort to \emph{ad hoc} hardware extensions and, as such, they cannot be used with off-the-shelf 
PM-based systems. 

After a decade of research based on performance models that could not be validated,
PM has finally reached commercial availability with Intel Optane™ Data Center Persistent Memory modules (\optane).
This constitutes a notable opportunity to revisit the research avenue paved by previous proposals, 
now with the ambitious goal of devising practical proposals whose design, implementation and evaluation is
no longer haunted by the above-mentioned limitations.




This paper is the first, to  our knowledge, to address that goal.

\textbf{As a first contribution}, we revisit the design space of a large body of pre-\optane state-of-the-art proposals and confront them with the idiosyncrasies of today's off-the-shelf \optane-equipped architectures.
This empirical study reveals that some key design choices in the literature rely on 
fundamental assumptions that are not verified in contemporary DRAM-\optane HMHs.
On one extreme, we find that simplistic placement policies that place read-only pages in PM and read-write pages in DRAM can
be highly inefficient when used with \optane. 
On the other extreme, we find that
sophisticated placement policies that balance pages across 
PM and DRAM to maximize the aggregate access bandwidth, when used with \optane, cannot achieve the gains that were foreseen in
pre-\optane literature. 
In-between both extremes, we focus on the \emph{fill DRAM first} placement policy and show that, to increase its potential advantages with \optane, both per-page hotness and
read-write intensity should be monitored and taken into account to guide the policy. \jpb{The last part of this paragraph is new. Please revise.}

From this study, we devise a set of design guidelines for tiered page placement that
take into account the lessons learned from this first part.

\textbf{As a second contribution}, we 
propose \name, a practical tool for tiered page placement in \emph{off-the-shelf} Linux-based systems equipped with DRAM+\optane HMHs.
In contrast to previous proposals, the design and implementation of \name follow a pragmatic approach that, 
from the ground up, is guided by two main practicality principles:
1) to be tailored to the performance idiosyncrasies of off-the-shelf DRAM+\optane systems;
and 2) to fit seamlessly into Linux with minimal kernel-mode components.

 In a nutshell, \name considers the disparity in performance between DRAM and \optane, and dynamically decides new page distributions that ultimately lead to a higher application throughput. 
 In order to achieve this, \name periodically identifies frequently modified and referenced pages, which benefit the most from the fast, but limited, DRAM tier.
 As a result, a larger portion of the workload's accesses is fulfilled by the DRAM tier, while \optane serves sporadically-read pages.
 We focus our efforts on a solution that requires minimal changes to the Linux kernel, and expands existing page placement mechanisms in order to accommodate the integration of \optane. \name is implemented into a contemporary Linux system by changing only 1 line in Linux' kernel code.
\rr{This sounds sort of obvious, is there something more insightful that can be highlighted? Perhaps say that the key pillars are not surprising, but applying them to this setting required X and Y.}\jpb{I agree. After I revise the architecture section, I'll come back to summarize the main design insights here. TO DO.}

 \textbf{As a third contribution}, 
 we present a comprehensive experimental evaluation of state-of-the-art proposals for tiered page placement, with
 realistic data-intensive workloads from NAS Parallel Benchmark (NPB) \cite{npb_article} and GAP suites \cite{gap_article}.
 To our knowledge, this is the first experimental study to 
 evaluate previous proposals on a real system equipped with \optane memory.
 Our results show that, in general, \name substantially outperforms previous state-of-the-art solutions with an average speedup of 4.6x in large footprint workloads.
 When compared to today's Linux default policy for tiered page placement in DRAM+\optane systems, \name achieves speedups of up to 11x. Considerable gains are also found in energy consumption.

The remainder of this paper is organized as follows. Section \Cref{section:background} provides background on \optane and data placement efforts proposed in past literature. \Cref{section:insights} studies the performance of a real \optane-equipped system, providing insights on how page placement should be attuned to such systems. \Cref{section:implementation} describes \name, detailing its design and implementation choices. \Cref{section:results} evaluates \name against existing tiered page placement alternatives.
Finally, \Cref{section:conclusion} draws conclusions.
\section{Background}
\label{section:background}

In this section, we provide background on tiered page placement for DRAM+\optane systems.
The section is organized in a bottom-up fashion.
In Section \ref{sec:bg:internals}, we start by describing \optane technology and how it is integrated into contemporary architectures.
Then, in Section \ref{sec:default_modes} we explain how \optane is exposed to the system software stack, and how the Linux operating system (OS) currently supports \optane.
Finally, in Section \ref{sec:relwork} we survey existing proposals to data placement 
in heterogeneous memory systems, starting by a broad perspective and then focusing on \optane.

T he main goal of our paper is to 
efficiently enable applications to place large \emph{volatile}
data sets in DRAM+\optane HMHs.
Therefore, our background is restricted to this scope.
As such, our presentation does not cover the use of \optane and other PM technologies to store crash-consistent persistent data structures, which is a complementary research avenue that has been surveyed in other papers \cite{10.1145/3465402,8642287}.
\rr{Mention that this discussion does not cover persistence, which as we mentioned is not relevant to the goals of this paper.} \jpb{Done (above).}


\subsection{\optane internals}
\label{sec:bg:internals}

\optane is delivered as DIMMs that are compatible with DDR4 sockets. 
The current capacity of \optane modules ranges from 128GB to 512GB, which represents up to a 4x increase in per-module capacity compared to DDR4 DRAM.
Currently, \optane modules can be used with 2nd or 3rd-generation Intel Xeon Scalable (Cascade Lake) CPUs.

Each CPU contains 2 integrated memory controllers (iMC), with 3 memory channels per iMC.
Each memory channel can have up to 2 DIMM slots (depending on the chosen motherboard), with the restriction that 
at most one \optane DIMM may be installed at a given channel. Therefore, a given memory channel can serve DRAM (up to 2 DIMMs), \optane (at most 1 DIMM) or a combination of 1 DRAM + 1 \optane DIMMs \cite{intel_dcpmm_manual}.
This flexibility enables these systems to have DRAM+\optane HMHs with varying DRAM-\optane capacity and bandwidth ratios.

Each iMC uses the DDR-T protocol to communicate with \optane. Like DDR4, DDR-T operates at cache-line granularity (usually 64B). 
Internally, each \optane module caches 256B blocks (called XPLines), with an associated prefetcher \cite{optane_study}. 
This cache also serves as a write-combining buffer for adjacent stores. 
Due to the granularity mismatch between DDR4 and XPLines, random stores incur in costly read-modify-write cycles \cite{optane_study4}.
Similar to SSDs, \optane uses logical addressing for minimizing wear-leveling, leveraging an internal address indirection table \cite{optane_study}. 



\begin{figure}[t]
\centering
  \begin{subfigure}{.45\linewidth}
    \centering
    \includegraphics[width=.99\linewidth]{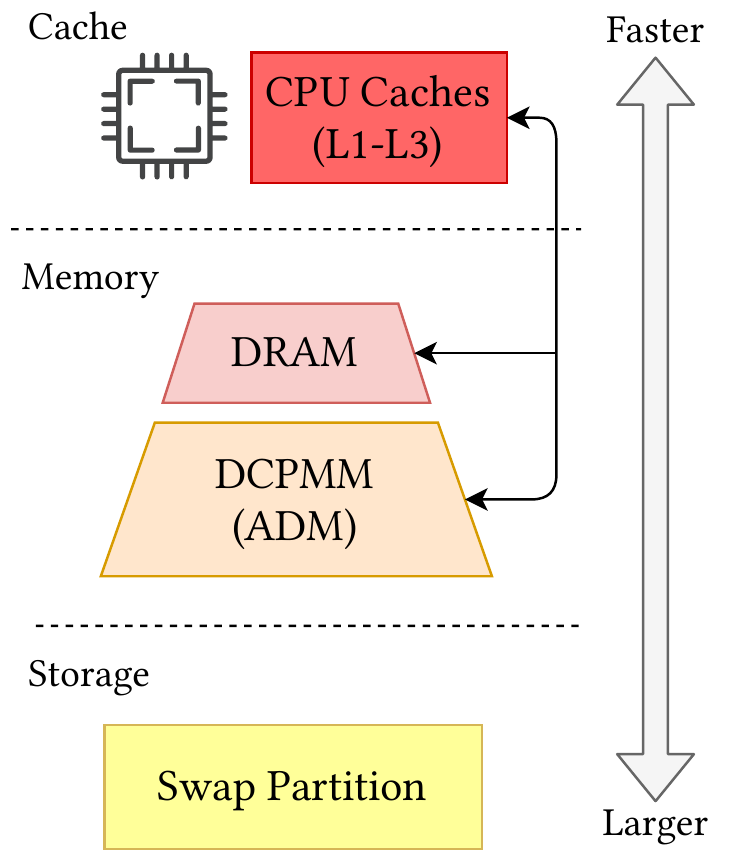}
    \caption{ADM Configuration}
    \label{subfig:optane-adm-arch}
  \end{subfigure}
  \begin{subfigure}{.45\linewidth}
    \centering
    \includegraphics[width=.99\linewidth]{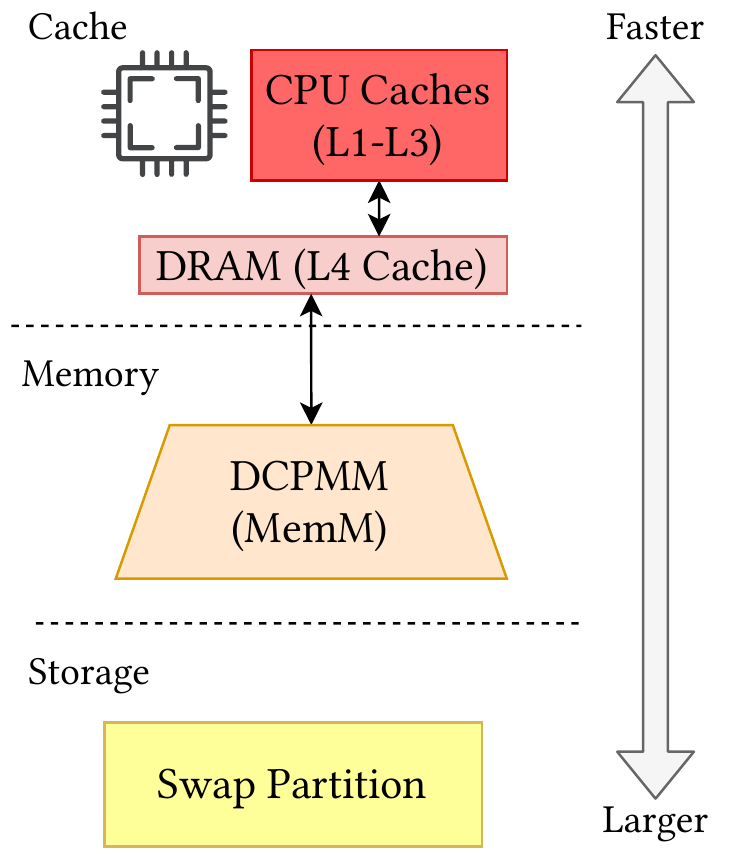}
    \caption{MemM Configuration}
    \label{subfig:optane-mm-arch}
  \end{subfigure}
   
  \caption[]{Main operation modes of \optane.}
  \label{fig:ambix_optane-config}
\end{figure}

\subsection{\optane operation modes and OS support}
\label{sec:default_modes}

\optane can be configured in two different modes: \emph{App Direct Mode} (ADM), and \emph{Memory Mode} (MemM), which \Cref{fig:ambix_optane-config} illustrates in a single-socket architecture. 
For simplicity, in the following we focus on a single-socket scenario.

In the MemM configuration, the \optane memory is transparently accessible to the OS as a single memory node, whose capacity is the same as the total \optane capacity installed in the socket.
DRAM is used as an internal last-level cache, which interposes every access to the local \optane memory node.

When \optane is configured in ADM, the installed DRAM and \optane memory are exposed to the OS as
distinct physical memory devices. Starting with Linux 4.0, both memory types can be abstracted as two distinct non-uniform memory access (NUMA) nodes.
The (virtual) pages of an application can be placed at either node and are directly accessed through load and store operations.
Therefore, applications have access to a larger aggregate main memory capacity, since the DRAM capacity is no longer hidden as a cache. 
In this setting, the default NUMA placement policy of Linux dictates that once a page is \emph{first-touched} it is placed on the fastest node (DRAM) as long as it has free space; otherwise, the slowest node (\optane) is selected. 
As with any NUMA system, Linux's \emph{numactl/libnuma} interface \cite{numactl} can be used to implement different policies.

\subsection{Related work}
\label{sec:relwork}

\subsection{Heterogeneous memory hierarchies (HMHs)} 
Historically, data placement in traditional HMHs (such as DRAM, SSD, disk) has relied on
inclusive caching policies. In such policies, any block from a lower tier needs to be copied to the top tier to be directly accessed from there. \rr{not sure if we need to explain what is an inclusive cache policy is?}\jpb{added.}
These are well-studied and a good fit for memory hierarchies whose 
adjacent layers differ in cost, capacity and performance by several orders of magnitude,
and where only the fastest layer can be accessed directly by applications \cite{Belady66}.

However, the appearance of new memory technologies has brought in new HMHs that invalidate these fundamental assumptions and call for radically different approaches to data placement.
Examples include the NUMA architectures that arise in multi-socket systems \cite{carrefour,asymsched}, 
high-bandwidth memory+DRAM memory hierarchies \cite{heteroos}, CPU-GPU memory hierarchies \cite{10.1145/3307650.3322224}, or software-defined far memories \cite{10.1145/3297858.3304053}, among others \cite{chopt}.
In such memory hierarchies, the traditional caching principle that each requested block needs to be brought into and 
served from the fastest memory becomes obsolete. 
In fact, Zhang et al. \cite{chopt} have shown that optimal data placement policies resort to
so-called \emph{cache bypassing} strategies \cite{jlpea6020005} that intentionally place some blocks in a slower memory to be accessed directly by applications.

\subsection{DRAM+PM memory hierarchies} 
More recently, DRAM+PM memory hierarchies have emerged as an additional case of a non-traditional HMH.
Triggered by the initial promises of the rise of PM technologies -- such as phase-change memory (PCM), Resistive random access memory (ReRAM), among others \cite{10.1145/3131848} --,
several works have proposed data placement techniques for DRAM+PM systems.
While, in theory, the main design principles of data placement in the above-mentioned memory hierarchies also apply to DRAM+PM systems, 
PM has distinctive performance and power characteristics, notably the  pronounced read-write access performance asymmetry~\cite{optane_study, optane_study2}.

To address the problem of data placement in DRAM+PM systems, some authors proposed programming or compile-time approaches. A first category is handcrafted applications which, using programming interfaces such as libvmmalloc library from the Persistent
Memory Development Kit (PMDK) \cite{scarg20book}, 
explicitly allocate each object on the most appropriate tier \cite{weiland_optane,10.1145/3447818.3460356,10.1145/3447818.3460355}. Alternative schemes resort to profiling-based code analysis to 
either guide \cite{x-mem,10.1145/3205289.3205320,8665751} or even automate \cite{auto_pena,unimem,membrain,10.1145/3368826.3377922} object allocation decisions. Alternatively, runtime systems for specific programming languages have bee proposed to manage object placement \cite{10.5555/3291656.3291698}.

The scope of our paper is on dynamic approaches to data placement for DRAM+PM.
Dynamic approaches typically manage data placement on a page granularity, therefore we hereafter designate them as \emph{tiered page placement} approaches.
In contrast to static compile-time approaches, dynamic page placement typically incurs no
programming burden, easily supports legacy applications, can dynamically adapt to workload changes by migrating
pages across tiers, and naturally manages multiple concurrent applications from a system-wide perspective.

\begin{table*}[!htb]
\tiny
\centering
 \begin{tabulary}{\linewidth}{| C | C | C | C | C | C | C | C |} 
 \hline
 \textbf{Proposed system} &
 \textbf{HMH assumptions} &
 \textbf{Page placement policy} & 
\textbf{Page selection criteria} &
\textbf{Page selection algorithm} & 
 \textbf{Modifications} &
 \textbf{Full-fledged implementation} &
 \textbf{Evaluated on \optane} \\
 \hline\hline

CLOCK-DWF \cite{clock_dwf} & DRAM+PCM & Partitioned & Hotness+r/w & CLOCK & OS &  &  \\
\hline
M-CLOCK \cite{mclock} & DRAM+PCM & Fill DRAM first & Hotness+r/w & CLOCK & OS &  &  \\
\hline
AC-CLOCK \cite{ac_clock} & DRAM+PCM & Fill DRAM first & Hotness+r/w & CLOCK & HW+OS &  &  \\
\hline
AIMR \cite{aimr} & DRAM+PCM/ReRAM & Fill DRAM first & Hotness+r/w & CLOCK+LRU & HW+OS &  &  \\
\hline
CLOCK-HM \cite{clock_hm} & DRAM+PCM & Fill DRAM first & Hotness+r/w & CLOCK+LRU & HW+OS &  &  \\
\hline
Seok et al. \cite{seok_second} & DRAM+PCM & Fill DRAM first & Hotness+r/w & LRU & HW+OS &  &  \\
\hline
DualStack \cite{dualstack} & DRAM+PCM & Fill DRAM first & Hotness+r/w & LRU & HW+OS &  &  \\
\hline
HeteroOS \cite{heteroos}, Nimble \cite{nimble} & MCDRAM+DRAM+NVM & Fill DRAM first & Hotness & LRU & OS & \cmark &  \\
\hline
UIMigrate \cite{uimigrate} & DRAM+PCM & Fill DRAM first & Hotness & LRU & HW+OS &  &  \\
\hline
TwoLRU \cite{twolru} & DRAM+PCM & Fill DRAM first & Hotness+r/w & LRU & HW+OS &  &  \\
\hline
Tiered AutoNUMA \cite{autonuma_optane} & DRAM+\optane & Fill DRAM first & Hotness+r/w & LRU & OS & \cmark & (\cmark) \\
\hline
Thermostat \cite{thermostat} & DRAM+3D XPoint & Fill DRAM first & Hotness & TLB misses & OS &  \cmark &  \\
\hline
Memos \cite{memos} & DRAM+NVM & Fill DRAM first + bandwidth balance & Hotness & TLB misses+CLOCK & OS & \cmark &  \\
\hline
Yu et al. \cite{yu_bw-hma} & DRAM-PCM & Bandwidth balance & n/a & n/a &  &  &  \\
\hline\hline
\textbf{\name} & DRAM-\optane & Fill DRAM first & Hotness+r/w & CLOCK+PCMon \cite{pcmon} & OS (1 line) & \cmark & \cmark \\
 \hline
\end{tabulary}
\caption{Comparison of proposals for tiered page placement on DRAM+PM systems.}
\label{tab:related}
\end{table*}

Table \ref{tab:related} summarizes the main proposals for tiered page placement in DRAM+PM systems and the respective design choices. 
The majority of these proposals precedes the commercial availability of PM, with \optane.
Therefore, they are implemented and evaluated on simulated or emulated environments, 
in most cases calibrated according to performance predictions for PCM PM \cite{pcm,pcm2,pcm3,pcm4,pcm5}, 
which is not the technology used in \optane.
Furthermore, many proposals simply overlook any integration into a real system stack --
they either focus on the algorithms and dismiss how they can be implemented, or resort to \emph{ad hoc} hardware extensions, making it impossible to implement them on off-the-shelf PM-based systems. 

We can categorize three main placement policies among the proposals listed in Table \ref{tab:related}.
Some adopt a \emph{partitioned} placement policy, which dynamically classifies each page as DRAM-bound or PM-bound according to
simplistic criteria that merely depend on each page's recent access history.
Such classification then dictates at which tier each page resides (with some exceptions, e.g., if DRAM is full).
A common such criterion is to classify read-dominated pages as PM-bound,
motivated by a simplistic assumption that the read performance of PM is comparable to DRAM. In this policy, pages are then migrated to DRAM when writes are detected. Proposals such as CLOCK-DWF \cite{clock_dwf}.

An alternative policy tries to \emph{fill DRAM first} as pages are accessed and, once a given DRAM occupancy threshold is 
reached, selects a set of pages to be spilled to PM. 
As Table \ref{tab:related} shows, this is the most prevalent policy among the existing proposals.
To select which pages to spill, some proposals use coldness as the single criterion, which conveniently allows them to reuse existing traditional cache replacement policies, such as LRU or CLOCK.
For instance, HeteroOS \cite{heteroos} and Nimble \cite{nimble} leverage the active and inactive page lists that Linux maintains concerning each NUMA node, and exploits them to implement tiered page placement. The coldness criterion can also be combined with read and write intensity, motivated by the read/write performance asymmetry of PM. This combination prioritize read-intensive cold pages over write-intensive ones as candidates to demote to PM \cite{mclock,ac_clock,aimr,clock_hm,seok_second,dualstack,twolru,autonuma_optane}.

Finally, a third placement policy is based on the observation that, with a pure \emph{fill DRAM first} strategy, bandwidth-intensive 
workloads might saturate DRAM bandwidth while not taking advantage of the available PM bandwidth.
Hence, a \emph{bandwidth balance} strategy tries to distribute hot pages across DRAM and PM, 
in some appropriate ratio, with the goal of maximizing the aggregate bandwidth that applications can attain when accessing different pages in parallel \cite{yu_bw-hma,memos}.

At the core of the above placement policies lie some mechanism to track page accesses, whose output
guides the placement decisions.
In recent years, different techniques have been proposed to tracking page accesses in DRAM-PM page placement systems, ranging from (i) monitoring page table access bits as in traditional page replacement algorithms (e.g. \cite{heteroos}), (ii) possibly complemented with TLB miss interception (e.g., \cite{thermostat,memos}), to
(iii) hardware extensions that track additional per-page metrics \cite{ac_clock,aimr,clock_hm,seok_second,dualstack,uimigrate,twolru}.

Their accuracy vs.\ runtime overhead trade-offs differs substantially, as previously authored noted (e.g., \cite{memos,thermostat}). 
Furthermore, different techniques require different levels of modifications to the underlying system stack, from incremental changes to the OS kernel (i) or deeper changes (ii), to modified hardware (iii). 
Some recent proposals \cite{thermostat,memos} resort to hierarchical approaches, which apply different tracking techniques  at different levels of granularity (from contiguous sets of pages to individual pages) to improve high tracking accuracy at low cost.

Finally, another core component of these solutions is the page migration mechanism.
Although Linux has native support for page migration across DRAM and PM nodes, 
some works have proposed optimized migration implementations.
These improvements are achieved by reducing the page management costs associated with migration \cite{10.1145/3296957.3173198,nimble}, 
exploiting bandwidth-optimized alternatives to copy page contents across memory nodes \cite{memos,nimble} (namely, exploiting multi-threaded or DMA-based data movement),
or providing new page exchange primitives \cite{nimble}.




\subsection{Tiered page placement with \optane} 
Since the availability of \optane, there are some initial studies of data placement in real DRAM+\optane systems.
Some authors studied the system performance of \optane \cite{optane_study,optane_study2,optane_study3,optane_study4}, and
others have manually modified data placement in either mini or real applications and evaluated the performance benefits in DRAM+\optane systems \cite{optane_study5,weiland_optane,10.1145/3447818.3460356,10.1145/3447818.3460355,10.1145/3357526.3357528,DBLP:conf/ics/XieDLPM021}.
Finally, recent profiling-based proposals to guide or automate static data placement have already been implemented and evaluated with real \optane \cite{10.1145/3368826.3377922,9139816}.
Still, with respect to \emph{dynamic} tiered page placement, to our knowledge, no previous work has systematically studied it on real \optane-based
systems with realistic applications (the only exception being the tiered Autonuma extension \cite{autonuma_optane}, whose documentation includes 
a brief evaluation on a synthetic benchmark). That is the goal of the following sections. \rr{If the last system already evaluated data placement, doesn't this contradict the claim that no previous work studied data placement in real systems?}\jpb{I revised this to highlight that nobody did this for dynamic page placement (different than static profile-based, or handcrafted)}

\section{Insights from real DRAM+\optane systems}
\label{section:insights}

This section revisits the main design choices of previous proposals for tiered page placement in DRAM+PM HMAs, in light of our experimental validation of some of the 
assumptions that underlie such choices.
Using a real machine equipped with \optane (see Section \ref{section:results} for detailed specifications), we conduct a set of 
focused experiments with the purpose of questioning whether each assumption, and the design choices that depend on it, holds
with \optane.
As we show next, this empirical evidence provides us with important insights concerning which design choices are most effective in practice, and which ones can or should be avoided.






A tiered page placement policy determines, at each instant, what is the set of pages that resides at each memory tier.
In this context, there are several factors that influence the overall performance and therefore serve as input to these policies.
Most notably, state-of-the-art proposals measure or estimate per-page hotness and per-page read/write intensity.

We start by studying
how these specific factors affect the performance of
different tiered page placement policies.
In this initial study, we consider a data set, which we divide into inactive and active pages.
Active pages are accessed by a multi-threaded application (with as
many threads as the available hardware threads), where threads perform sequential%
\footnote{Due to space limitations, we omit random-access workloads. We note that switching to random-access workloads amplifies the 
per-access costs (when comparing \optane to DRAM)  that we present next.} 
accesses to non-overlapping regions. Inactive pages, in turn, are never accessed during our experiments.
For each test, we vary two main dimensions of the workload: i) the access demand, which determines the hotness of the active pages, by varying the time each thread stalls between
memory accesses; and ii) the read/write ratio, ranging from \emph{all reads} to a 2:1 read-write ratio.
We evaluate each workload with two static placement configurations: i) all active pages in DRAM, ii) all active pages \optane. 

We instantiate the above tests using the Intel Memory Latency Checker (MLC) benchmark \cite{mlc} on our \optane-equipped machine. 
MLC measures read latency and bandwidth of the different tested scenarios.

\begin{figure}[t]
  \centering
  \includegraphics[width=0.48\textwidth]{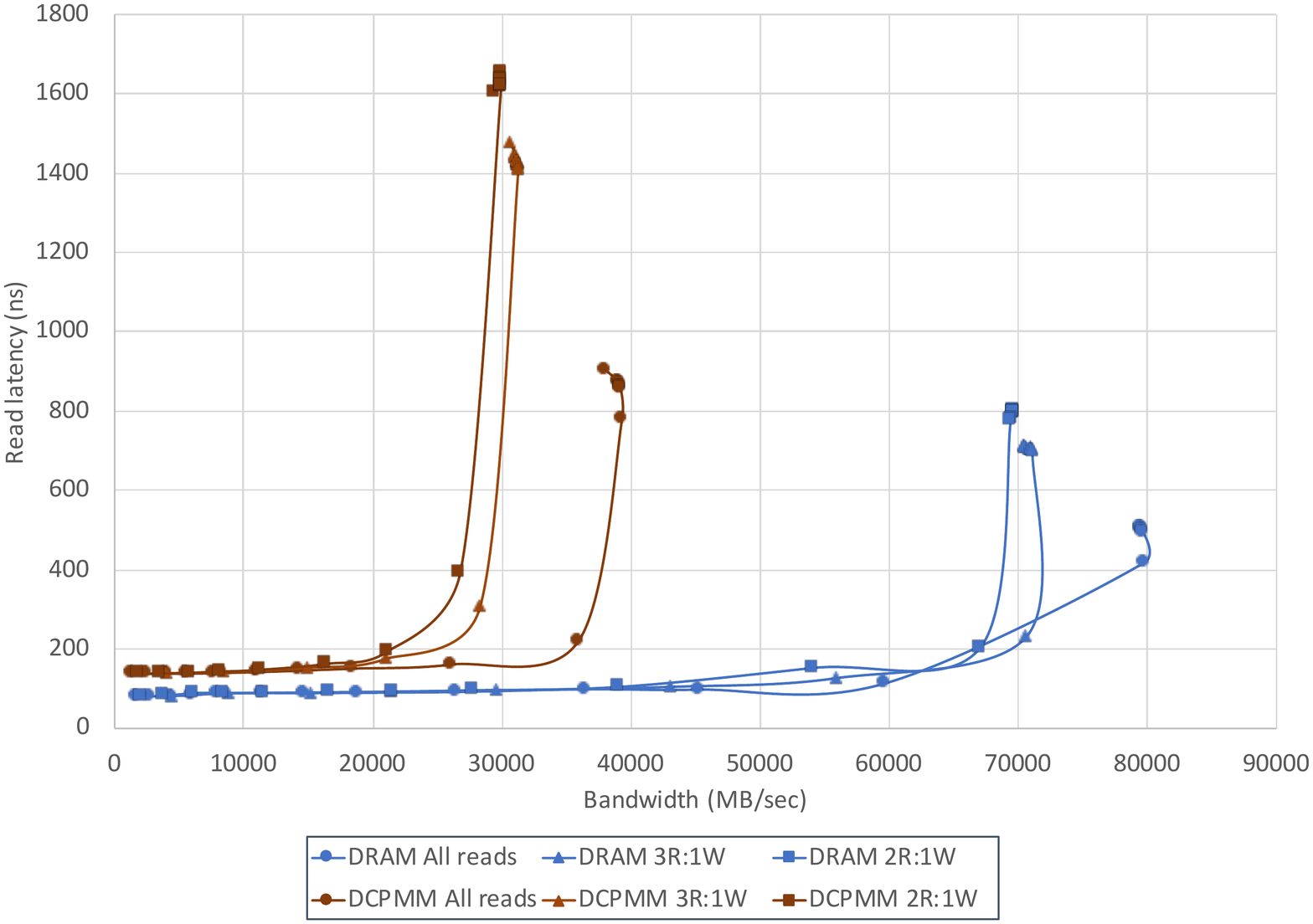}
  \caption{Latency and bandwidth for DRAM and \optane, for different read/write intensities (different lines) and memory access demands (different points in each line).}
  \label{fig:latency-profiles}
\end{figure}


Figure \ref{fig:latency-profiles} 
presents the results.
Each line represents a workload with a given read/write ratio, and each point in a line denotes an increasing access demand level.
The set of lines corresponding to a given memory tier delimit its performance lower and upper bounds 
(concerning latency and bandwidth). 
We refer to the interval defined by each set of lines as the DRAM or \optane regions.

This figure therefore allows us to analyze different placement policies, by interpreting them as follows. 
Let us assume that, at some instant, a given application (or system)
has a set of pages in DRAM and another set in \optane.
Given each set, let us consider the overall access demand and read/write intensity of the accesses that target the pages in that set (i.e., in the corresponding memory tier).
For each tier, Figure \ref{fig:latency-profiles} 
enables us to identify a point that corresponds to such access demand and read/write intensity in the associated tier's region, thus obtaining the corresponding read latency and bandwidth.

\subsection{\emph{Partitioned} policy.} As a first policy that we analyze, consider a \emph{partitioned} page placement scheme, which places every read-dominated pages in \optane and the remaining pages in DRAM (as in \cite{clock_dwf}).
In this case, the \optane point is located somewhere in the \emph{all reads} line of \optane, depending on the access demand to such pages. 
We observe, in this case, that with workloads that have a small subset of read-write active pages, a considerable portion of DRAM space will be unused.
Such space would be used more efficiently if it served a fraction of the active read-only pages, with the associated performance gains.
By comparing the \emph{all reads} curve of \optane with its DRAM counterpart, we see that, by not placing read-only pages in free DRAM space, the \emph{partitioned} policy can lead to up to 11.3x latency costs and up to a 2x drop in peak bandwidth (when such pages are concerned).\jpb{I updated the sentence above with the concrete values; I also changed the observation below to be more abstract (I moved the figures to the text that precedes each highlighted observation)}

\vspace{2mm}\noindent\textbf{Observation 1}: for workloads whose set of read-write active pages is smaller than DRAM, the \emph{partitioned} policy can incur substantial latency and bandwidth costs.
\vspace{2mm}

\subsection{\emph{Fill DRAM first} policy.} In contrast, a \emph{fill DRAM first} policy is able to mitigate the above
inefficiency. Hence, such policy is more suitable to real \optane-based
systems.
As discussed in Section \ref{sec:relwork}, while some existing proposals exclusively guide their placement decisions by the page hotness metric, more sophisticated designs
combine hotness with per-page read/write intensity.
While more elaborate, the latter try to exploit the strong read vs. write access asymmetry of \optane.

By analyzing Figure \ref{fig:latency-profiles}, 
we can infer how much taking read/write intensity into account matters to tiered page placement. 
As an example, let us first consider that, with a policy that is agnostic of read/write intensity, 
the set of (hotter) active pages in DRAM has a similar read/write intensity as the set of (colder) active pages in \optane.
Further, let us assume that the average read/write ratio of both sets is 2R:1W.
Hence, the access performance to the active pages in each set is represented by some point in the 
2R:1W line of the corresponding tier in Figure \ref{fig:latency-profiles} (the actual location of such points
depends on the access demand that the application places on each set of pages).
If, instead, we consider a policy that
favors write-intensive pages in DRAM and read-intensive pages in \optane,
this change can be visualized as moving the DRAM point upwards (to a more
write-intensive mix of pages) and moving the \optane point downwards (to a more read-dominated mix). 

Figure \ref{fig:latency-profiles} enables us to quantify the potential impact of this approach.
While access demand is low (i.e., the first points of each curve), the different lines that characterize 
each tier are relatively overlapping; therefore, for low access demand, a read/write-aware policy is not expected
to yield evident performance gains. 
Still, as we increase access demand to higher levels, the \optane curves 
start diverging substantially after the 20,000 MB/sec limit, 
exposing increasingly prominent read vs. write access asymmetries.
In contrast, the DRAM curves only express clear read vs. write asymmetries when stressed at extreme levels (beyond the 60,000 MB/sec limit), but always
with much smaller discrepancies than those observed with \optane.
Consequently, for workloads with medium to high bandwidth intensity, 
prioritizing write-intensive pages in DRAM introduces large performance gains when pages in \optane are accessed, 
while imposing residual to modest performance costs to DRAM accesses.
Hence, the net balance will, in many cases,
be positive by a considerable margin.\jpb{I reformulated this paragraph, please revise it.}

\vspace{2mm}\noindent\textbf{Observation 2:} For bandwidth-intensive workloads, considering read/write intensity when choosing which pages to migrate in a \emph{fill DRAM first} policy can lead to
substantial performance gains when compared only considering per-page hotness. 
\vspace{2mm}

\begin{figure}[t]
  \centering
  \includegraphics[width=0.48\textwidth]{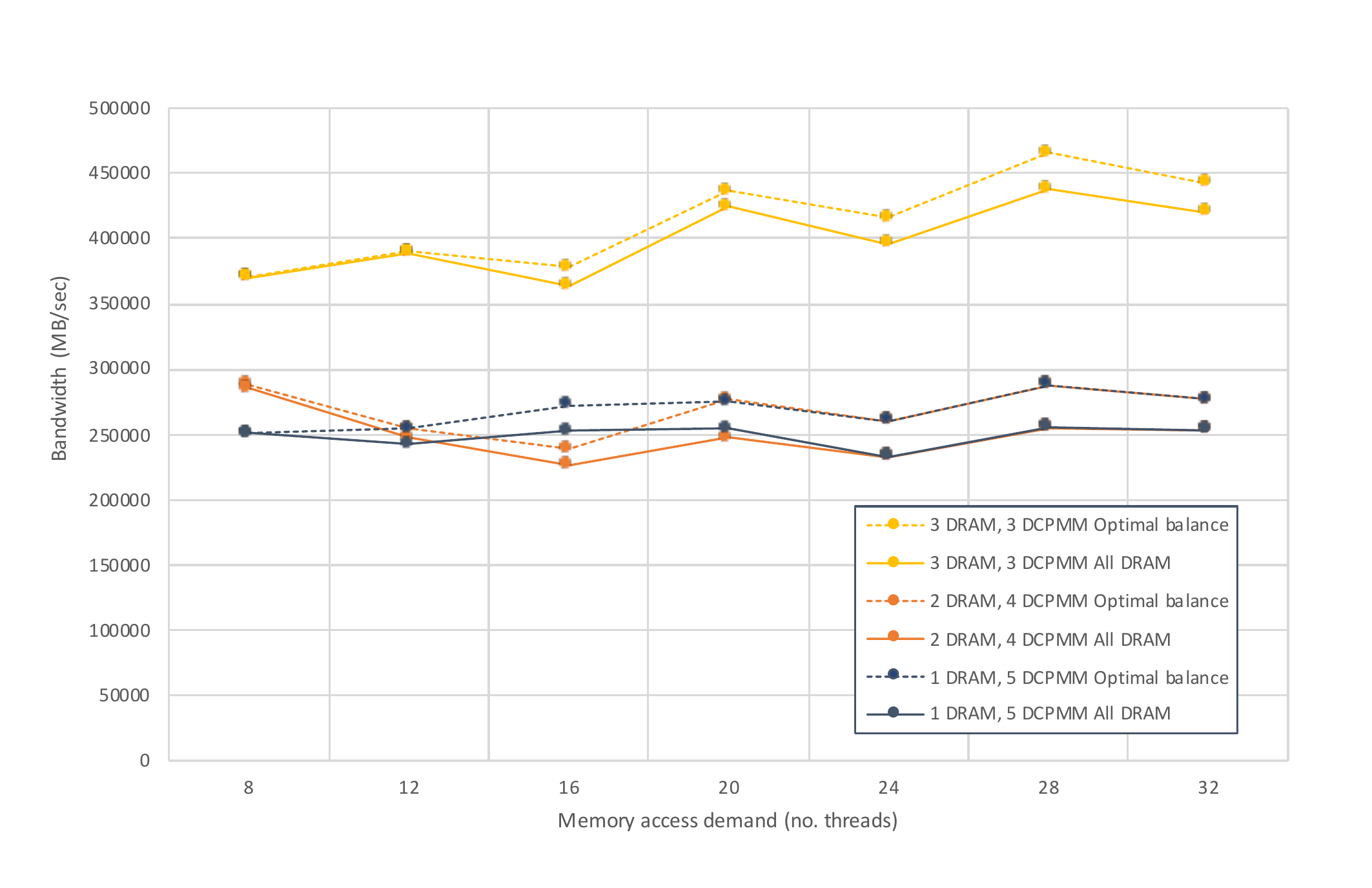}
  \caption{Effective bandwidth gains achievable by an ideal \emph{bandwidth balance} with read-only workloads of varying memory access demand levels, under different memory module configurations.}
  \label{fig:bw-balancer-potential}
\end{figure}

\subsection{\emph{Bandwidth balance} policy.} As a final question, we wish to quantify what are the
potential benefits of a \emph{bandwidth balance} policy
in contemporary \optane-based systems, when compared to a simpler \emph{fill DRAM first} approach.
A \emph{bandwidth balance} policy tries to balance active pages across DRAM and \optane such that
the application will distribute its concurrent accesses across both tiers, rather than concentrating its
activity on a single tier. 
The goal is that the application benefits from the \emph{aggregate} bandwidth that both tiers provide, which in theory is larger than the bandwidth of the DRAM tier only.

A noteworthy observation from the previous insight is that the additional bandwidth that \optane can contribute varies substantially, depending on the read/write intensity. 
In fact, for write-intense workloads, \optane bandwidth becomes so low that its contribution to the aggregate bandwidth becomes residual, as Figure \ref{fig:latency-profiles} shows.
Hence, in order to obtain relevant bandwidth gains, a \emph{bandwidth balance} must be able to select only read-dominated active pages to reside in \optane. 

Therefore, to quantify the potential advantage of an ideal \emph{bandwidth balance}, 
we return to our initial benchmark, and focus exclusively on the \emph{all reads} workload -- i.e., the best-case scenario for \emph{bandwidth balance}.
In contrast to the previous experiment, this time we distribute pages across
tiers according to different ratios (100\% in DRAM, 95\%, 90\%, ...), using weighted-interleaved placement \cite{bwap}.
Also, this time we vary the number of active threads to test different memory access rates.
For each access demand value, we select the DRAM/\optane distribution ratio that obtained the optimal performance (more precisely, that minimized execution time).

The experiments were performed on our \optane-based machine with 
all available memory channels populated. 
We considered different physical memory configurations, by varying the number of channels used by DRAM and \optane modules, from 3:3 (lower \optane bandwidth) to 1:5 (higher \optane bandwidth). 
Figure \ref{fig:bw-balancer-potential} presents our findings.

A first finding is that a balanced placement is only advantageous when bandwidth intensity is high enough.
Up to 8 (with 2:4 and 1:5) and 12 threads (with 3:3), the best performing configuration is to simply place every page in DRAM.
The effective advantage of \emph{bandwidth balance} only becomes noticeable at very high access demand rates. To understand this observation, recall from Figure \ref{fig:latency-profiles} that DRAM latency is much lower than \optane when DRAM bandwidth 
is not saturated. 
Therefore, the break-even point for \emph{bandwidth balance} only occurs with bandwidth-intensive workloads that are able to saturate DRAM bandwidth. 

Furthermore, even when we consider bandwidth-intensive workloads, the 
actual gains of \emph{bandwidth balance} turn out to be 
relatively disappointing, yielding modest speedups (at most 1.13x).
These results show that the actual aggregate bandwidth that \emph{bandwidth balance} may achieve is, in fact, much lower than
the sum of the nominal peak bandwidth of DRAM and \optane.

\vspace{2mm}\noindent\textbf{Observation 3}: The potential advantage of a \emph{bandwidth balance} policy is very limited in contemporary \optane-based systems.
\vspace{2mm}
\jpb{I revised the above discussion about bw balance, to reflect the new values in the plot. Please review it.}

It is worth remarking that, while our machine is populated with one DRAM or \optane module per channel, 
some motherboards support up to two modules per channel. However,
while\jcm{?}
at most one \optane module can be installed per channel, a fully populated 
channel will have an \optane module and a DRAM module competing for the channel's bandwidth, which further reduces the perceived \optane bandwidth, 
hence is not expected to change the above results. \jpb{this is new, please revise} 

\section{\name}
\label{section:implementation}

In this section we present \name, a system that makes page placement decisions based on the lessons from the previous section. 
We start by mapping those insights to the key design decisions, and then detail the design and implementation of the system.

\subsection{From insights to architecture}

Guided by the insights we draw in the previous section, we exclude from our design both the \emph{partitioned} and the \emph{bandwidth balance} policies (based on Observations 1 and 3). 
Instead, based on Observation 2, we adopt a \emph{fill DRAM first} policy that selects pages to
migrate based on a combination of per-page hotness and read/write intensity.


More precisely, \name collects page and bandwidth metrics at runtime in order to perform page placement decisions that take advantage of each tier's characteristics.
To this end, \name classifies pages into three different categories: write-intensive, read-intensive and cold.
Then, the algorithm keeps as many write-intensive pages as possible in DRAM.
If these do not fully occupy DRAM, \name then prefers read-intensive pages over cold pages in the faster tier.
As we will show, this simple set of design choices coupled
with a portable implementation that fit seamlessly into Linux with minimal kernel-mode components allows \name to be more efficient and practical than existing proposals. 

\subsection{System design}

\name manages page placement within a socket with a DRAM and a \optane (in ADM) tier.
In multi-socket systems, multiple instances of \name run independently, each managing its own socket.
For simplicity of presentation, in the following description we assume a single-socket scenario.
\jpb{I feel that this detailed remark interrupts what should first be a high-level overview. Still, I couldn't manage to push it to a later stage, since the remaining of the text assumes single socket. Any suggestions?} \rr{Restructured a bit, is it better this way?}

The design of \name is based on the key idea of having a set of properties about the page distribution that it tries to maintain, and then react by migrating pages only in case those target properties are not met. This target suitability of the current page distribution is defined by the following three criteria:
\begin{itemize}
    \item DRAM has enough free space to allow newly referenced pages to fit in the faster tier.
    These pages are expected to be accessed frequently after their allocation, due to the \emph{temporal locality} principle.
    Thus, \name maintains a defined buffer of free space in DRAM by demoting pages eagerly, before it is depleted.
    \item \optane's write throughput is nominal, indicating that the tier does not contain a significant amount of frequently modified pages.
    \item If DRAM is at capacity but the \optane's write threshold is substantial, then this is also considered to be on target, since no pages can be exchanged between both tiers, such that the threshold is reduced.
\end{itemize}

If the current distribution does not meet any of these criteria, \name devises a new placement decision that corrects it, migrating a subset of pages in a given orientation.

To select which pages to migrate, \name leverages the unmodified page table walk (pagewalk) and PTE bit manipulation mechanisms, implemented in the Linux kernel.
These mechanisms have a relatively stable implementation, which benefits \name by making it compatible with a wide range of kernel versions.
Furthermore, configuring \name on a new kernel version requires only a single line of code, which exports the pagewalk routine, making it available to our solution.

In demotion scenarios, we apply concepts from the traditional CLOCK algorithm, modified to separate intensive pages into read- and write-dominated.
Page promotion, on the other hand, applies a novel \comp{delay} mechanism, which allows \name to identify recently accessed and modified pages in the \optane tier with low overhead.

\name also implements an exchange-based migration technique, using only pre-existing system calls, wherein an equal number of pages are switched between both tiers, thus preserving their current allocation.

\subsection{Architecture}

\begin{figure}[t]
  \centering
  \includegraphics[width=0.4\textwidth]{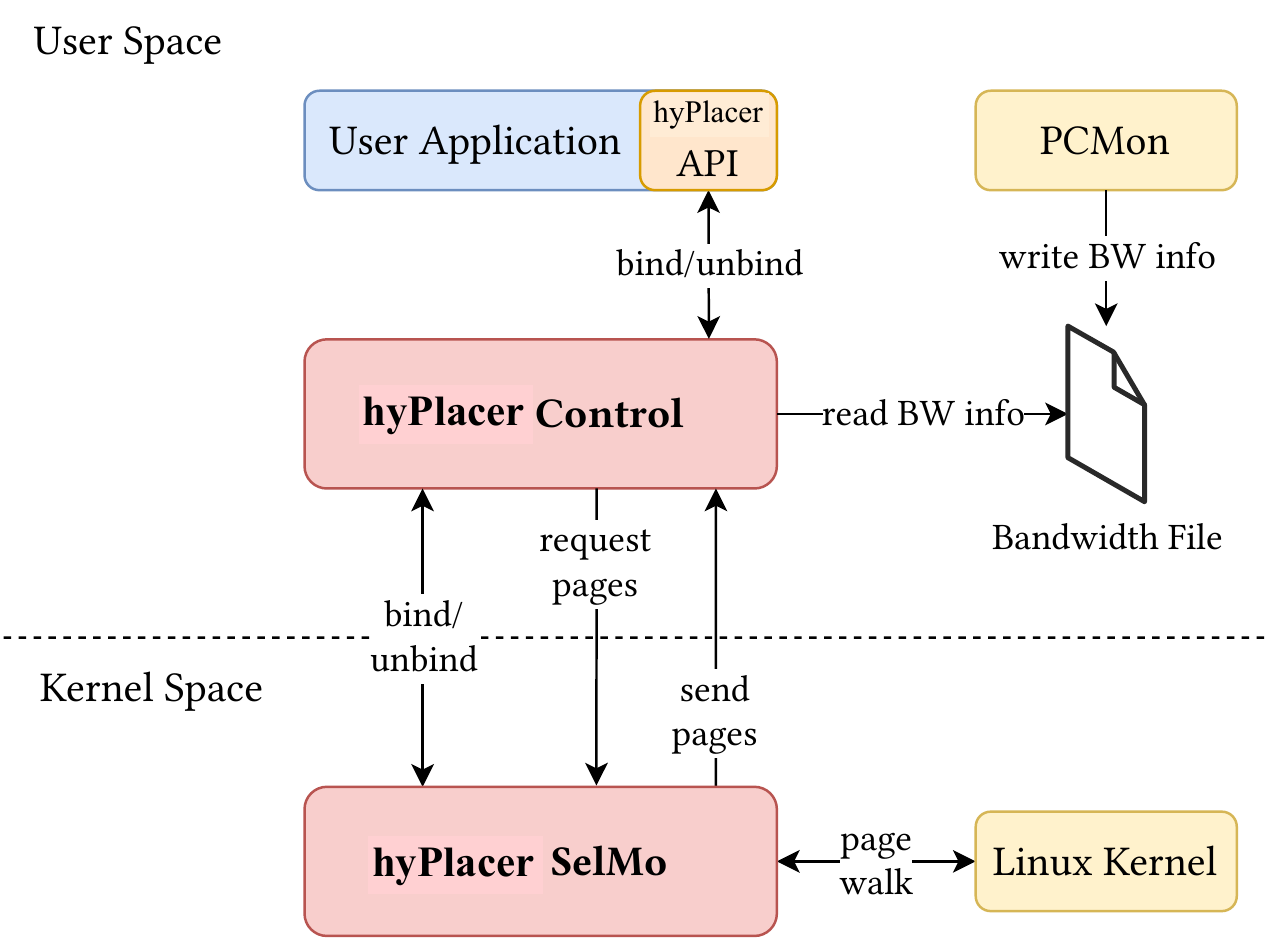}
  \caption{\name architecture overview.}
  \label{fig:architecture}
\end{figure}

\Cref{fig:architecture} provides an overview of \name, presenting its components and how they interact.
Our solution consists of two main components: \comp{Control} and the Page Selection Module (\comp{SelMo}).

\comp{Control} is an elevated process running in user-space, which is responsible for formulating and putting into effect new placement decisions, and binding/unbinding applications to our solution.
In order to formulate new decisions, the component leverages Processor Counter Monitor (\comp{PCMon})~\cite{pcmon}, which periodically outputs the current throughput per node to a shared text file.

To use kernel-implemented mechanisms, \name integrates a kernel module, named \comp{SelMo}.
The module selects pages belonging to bound processes, in order to carry out the \comp{Control}'s decisions.

The dual-component architecture achieves a small footprint in kernel-space, since all mechanisms related to devising and carrying out new placement decisions are offloaded to the user-space component.

\subsection{Implementation}

Overall, 
\name leverages existing page management mechanisms already supported in Linux: (i)~existing page walking mechanisms, (ii)~the page table's dirty and reference bits, managed by the memory management unit (MMU), (iii)~the \textit{move\_pages} syscall to migrate pages between tiers, and (iv) Processor Counter Monitor (PCMon) \cite{pcmon}, which allows \name to determine the bandwidth usage of each of the memories with hardware counters available in most modern Intel CPUs.

\comp{Control} periodically monitors current memory usage and throughput values per NUMA node. 
It is also able to detect the presence of write-intensive pages in \optane, without communicating with \comp{SelMo}, by reading the information file generated by \comp{PCMon}.
Depending on the collected metrics, \comp{Control} devises a new placement decision and sends a \comp{PageFind} request to \comp{SelMo}.
The request contains the number of pages to find and specifies the selection criteria, or mode.

\begin{table}[t]
\tiny
\centering
 \begin{tabular}{| c | c | p{4cm} |} 
 \hline
 \textbf{Mode} & \textbf{Tier Scope} & \multicolumn{1}{c|}{\textbf{Goal}} \\
 \hline\hline
 \comp{DEMOTE} & DRAM & Demote cold pages \\ 
 \hline
 \comp{PROMOTE} & \optane & Promote pages \\
 \hline
 \comp{PROMOTE\_INT} & \optane & Promote only intensive pages \\
 \hline
 \comp{SWITCH} & Bidirectional & Switch intensive with cold pages \\
 \hline
 \comp{DCPMM\_CLEAR} & \optane & Clear the R/D bits from all resident pages \\
 \hline
\end{tabular}
\caption{\label{tab:find} \comp{PageFind} modes and goal.}
\end{table}

In \Cref{tab:find} we summarize the multiple modes, describing the tier from which they select pages, and their respective goals.
By design, DRAM has a defined maximum usage threshold below its actual size.
Above the threshold, \name considers that the tier is full or near to depletion, and requests DRAM-resident cold pages to demote from \comp{SelMo}, via a \comp{DEMOTE PageFind}.

Similarly, \optane has a defined write throughput threshold.
If \optane's current throughput is above the defined threshold, \comp{Control} requests intensive pages and promotes them to DRAM.
If DRAM is above its usage threshold, an equal number of pages must be demoted, such that the free space buffer is preserved.
Therefore, a \comp{SWITCH PageFind} is sent to the module.
Otherwise, it tries to maximize the faster tier's utilization, by promoting as many intensive pages as possible such that the usage threshold is not surpassed, by requesting a \comp{PROMOTE\_INT PageFind}.

Inversely, if the \optane's throughput is below its threshold, two decisions can be performed.
If DRAM has enough available space, \comp{Control} allows cold pages to be eagerly promoted, via a \comp{PROMOTE PageFind}.
Otherwise, if DRAM is near depletion, a \comp{DEMOTE PageFind} is requested, such that cold pages are demoted.

Before sending any \comp{PageFind} request that will promote pages, \comp{Control} requests \comp{SelMo} to clear the R/D bits of all PTEs pointing to pages in the \optane tier, via \comp{DCPMM\_CLEAR PageFind}, after which it waits for a configurable \comp{delay}.
\comp{Delay} affects the access frequency at which a page is considered intensive.
Pages that are accessed or modified during the \comp{delay} interval are considered read- or write-intensive, while all others are classified as cold.

After receiving a \comp{PageFind} request, \comp{SelMo} iterates over each bound process' page table, in order to select which pages \comp{Control} should migrate.
The module leverages the kernel-implemented routine \code{walk\_page\_range()} , which iterates over a defined virtual address range.
Although this routine is not originally invocable from arbitrary kernel modules, 
we expose it to kernel modules by changing a single line of code in the Linux kernel. This is the only change to kernel code that \name requires. 

\comp{SelMo} passes a PTE callback as an argument to the routine, so that it can observe and manipulate each PTE's R/D bits. 
Since we want to perform different operations depending on the goal of the \comp{PageFind} request, we define multiple callbacks, one for each mode.

If the callback's goal is to demote pages, then it clears the R/D bits of all pages that are not selected to demote.
If one such page is referenced thereafter, the memory management unit (MMU) sets its PTE's reference bit; and also its modified bit, in the event of a store operation.
In contrast, if the page is not accessed until the next page table iteration, then it is suitable for demotion.

If, otherwise, the callback's goal is to promote pages, then PTEs are expected to have both their R/D bits unset, since they have been recently cleared by the module.
In this scenario, the MMU may change the PTE's R/D bits, so that the respective page is suitable for promotion over a next page table iteration.
Therefore, promotion callbacks do not directly manipulate R/D bits.
Instead, the algorithm deems a page in the \optane intensive if referenced during the \comp{delay} window, and write-dominated if modified.

When: (i) the number of selected pages exceeds the required amount set in the request; or (ii) the process has iterated over all PTEs, the page selection phase ends.
At this point, the last PTE's address and PID are stored and the page selection phase ends.
For each tier, the module keeps two last address and PID pairs, which set the start of the next page selection phase for that tier.
Thus, PTEs that have not been inspected for longer are prioritized for migration over recently seen ones.
Then, a reply-back phase begins, where \comp{SelMo} prepares a final page array to be sent back to \comp{Control}, containing all selected pages, after which the latter migrates them.




















\section{Evaluation}
\label{section:results}

In this section we evaluate the performance of \name when managing different realistic workloads and compare it against: (i) HMH-aware dynamic placement solutions recently proposed in literature, and (ii) placement options that are currently available in off-the-shelf \optane-equipped Linux systems.
In particular, our evaluation studies the throughput and energy consumption of each placement alternative with workloads of different sizes, read/write ratios, locality, access patterns and memory intensities.

Section \ref{sec:eval:setup} starts by detailing the experimental setup.
Section \ref{sec:bigdata} then presents the throughput and power results,
discussing how the different systems perform when managing realistic applications with large data sets. 
Finally, Section \ref{sec:eval:smalldata} analyzes the overheads of the different solutions when used in a worst-case scenario of small data sets. 

\subsection{Experimental setup}
\label{sec:eval:setup}

\noindent\textbf{Machine.} We use a dual-socket machine with Intel\textsuperscript{\tiny\textregistered} Xeon\textsuperscript{\tiny\textregistered} Gold 5218 CPU, running at 2.30GHz.
Each socket is populated with 2 DRAM modules (2x16GB DDR4 2666 MT/s) and 2 \optane modules (2x128 GB, DDR-T 2666MT/s Series 100), resulting in a total of of 32GB of DRAM and 256GB of \optane per socket. 
Each module is installed at a distinct memory channel. 

Unless otherwise noted, all experiments run on a Debian-based v5.8.5 kernel.
We set the \textit{swappiness} value to 0, which effectively disables swapping in our experiments. 
We restrict every experiment to a single socket, by using \emph{numactl} to bind each application to exclusively execute on 
the cores of the first socket and place its pages in the local DRAM and \optane nodes.
Finally, we disable transparent huge pages (THP), since we observed that the current native support in Linux for THP migration is very inefficient (since each THP is split and migrated as 4KB pages); the only exception is explained below.

\noindent\textbf{Evaluated solutions.} We evaluate a wide set of alternative approaches to tiered page placement.

On a first category, we consider the two available alternatives that are available by default in current Linux-based systems with \optane.
Namely, we consider: (a) DRAM and \optane configured in ADM, with the default \textit{first-touch} NUMA policy, without any dynamic placement solution applied (\textit{ADM-default}); and (b) \optane configured in MemM.
As explained in \Cref{section:background}, the former is a two-tiered configuration with no tier migration, whereas the latter provides a hardware-managed caching algorithm, which dynamically caches and evicts intensive data to and from DRAM.

On a second category, we evaluate a selection of recent proposals for tiered page placement from the literature.
As a selection criteria, we considered all the proposals in Table 
\ref{tab:related} that were designed and implemented to run on commodity systems.
This comprises: 

\begin{enumerate}
    \item \textit{autonuma}: Intel's extension of the original AutoNUMA NUMA scheduling mechanism of Linux \cite{autonuma_optane}, which adds tiered page placement to AutoNUMA's 
    original features. 
    Since we confine our experiments to a single-socket, \textit{autonuma}
    does not place or migrate threads or pages across sockets, which is its most common functionality. Instead, it simply manages tiered page placement between DRAM and \optane on a single socket.
    
    We used the \textit{tiering-0.4} version~\cite{autonuma_git}, which was the most up-to-date documented version at the time of testing, based on the v5.5 Linux kernel. We configure the kernel and run the post boot setup as proposed in the documentation, using the recommended settings for performance experiments.

    \item \textit{memos}, the tiered page placement of the Memos \cite{memos} OS kernel, an extension of the Linux kernel. 
    Memos employs an intricate adaptive and read/write-aware \emph{bandwidth balance} policy.
    Since Memos' code was not publicly available, we implemented the proposed page placement policy on top of the architecture of \name, relying on the mechanisms implemented in \name to monitor page hotness and read/write intensity.
    Note that we omit features proposed in \cite{memos} such bank imbalance, alternative migration techniques, and an in-house TLB miss profiler, as implementing these required deep kernel modifications. 
    
    Memos’ original parametrization \cite{memos} is suited to low
    footprint workloads, only migrating a maximum of 10,000 pages at each cycle, which correlates to 1MB/s.
    Therefore, we change it to make it as competitive as possible with our chosen workloads, as some will
    be multiple times larger than DRAM size (>27GB).
    Firstly, we tighten \textit{memos}' periodicity from 40s to 4s and, in order to fit in the new 4s period, lower the
    number of required page classifications to a single one, sacrificing accuracy for performance. Secondly,
    we increase the maximum number of pages that can be migrated in a given period to 10x its original
    value, allowing 100,000 pages to be promoted. Both changes increase \textit{memos}' migration rate-limit a
    hundredfold, to 100MB/s.

    \item \textit{nimble}, the tiered page placement of Nimble \cite{nimble}. This is an 
    incremental extension to the active and inactive page lists maintained by Linux for
    each NUMA memory node, which distinguish hot and cold pages, respectively. Using the
    same strategy originally proposed in HeteroOS \cite{heteroos}, Nimble uses such lists
    to implement a \emph{fill DRAM first} policy.
    We directly used the open source Nimble patch to the Linux v5.6rc6 kernel, with the default parameters described in the paper. Since Nimble also incorporates an optimized support for THP migration, we enabled it for the experiments with this system, in order to take advantage of its full feature set.

\end{enumerate}

Finally, a full-fledged implementation of \name was tested and compared against the above alternatives. We used the following main parameters: a DRAM occupancy threshold of 95\%; at most 128K pages migrated at each activation;
an \optane bandwidth threshold of 10 MB/sec; and an access bit clearance delay of 50 ms.


\noindent\textbf{Workloads.}
We evaluate multiple applications from the 
NAS Parallel Benchmark (NPB)~\cite{npb_article,npb_site} suite, extracted from the OpenMP~\cite{openmp} version of NPB v3.4.1.
NPB provides benchmarks that mimic common access patterns in computational fluid dynamics applications, and was designed to evaluate the performance of parallel supercomputers. Within the NPB suite, the selection criteria was to include the applications that could be instantiated with larger data sets (much larger than the available DRAM, 32 GB). 
As a result, we used the BT, FT, MG and CG applications. Except for the parameters affecting data set size, we used the default arguments.
All applications run with as many hardware threads as possible in the CPU, 32 threads.

Table \ref{tab:npb} summarizes the selected applications. 
Overall, the applications are characterized by different memory intensities, as well as read vs. write ratios. 

\begin{table}[t]
\tiny
\centering
 \begin{tabulary}{\linewidth}{| C | C | C | C | C | C |} 
 \hline
 \textbf{Benchmark} & \textbf{Read/write ratio} & \textbf{Data Sets Sizes (GB)} & \textbf{Average execution time} \\
 \hline
 BT
 & 3.5R:1W
 & 28.4 (S), 39.1 (M), 53.9 (L) &  9m (S), 16m (M), 45m (L)  \\
 \hline
 FT
 & 1.7R:1W & 20 (S), 40 (M), 80 (L) & 3m (S), 1h38m (M), 6h29m (L)  \\
 \hline
  MG
 & 4R:1W & 26.5 (S), 74.3 (M), 131 (L) & 6m (S), 1h16m (M), 2h53m (L) \\
 \hline
 CG
 & $>$60R:1W & 18 (S), 39.8 (M), 150 (L) & 12m (S), 41m (M), 3h58m (L) \\
 \hline
\end{tabulary}
\caption{\label{tab:npb} Summary of evaluated applications.\jpb{@Ilia it would be great to collect peak bw usage, when the application is running with the small data set, all in DRAM (David Gureya can help on this).} }

\igk{David have already collected this data for some workloads... methodology described, i can do it for the rest. I'm not sure I understand fully this table. It includes only one test per workload, whereas values would be dramatically different for different sizes, which one is considered?}\jpb{good question: for peak bw usage, I guess you can simply pick the value from the size that yields the highest bw usage (we need to clarify that in the text)}\igk{But it means we need to re-run all tests measuring maximum bandwidth and peack the maximum... Unfortunately haven't run it yesterday}
\end{table}
\begin{figure*}[t]
\centering
    \includegraphics[width=0.95\linewidth]{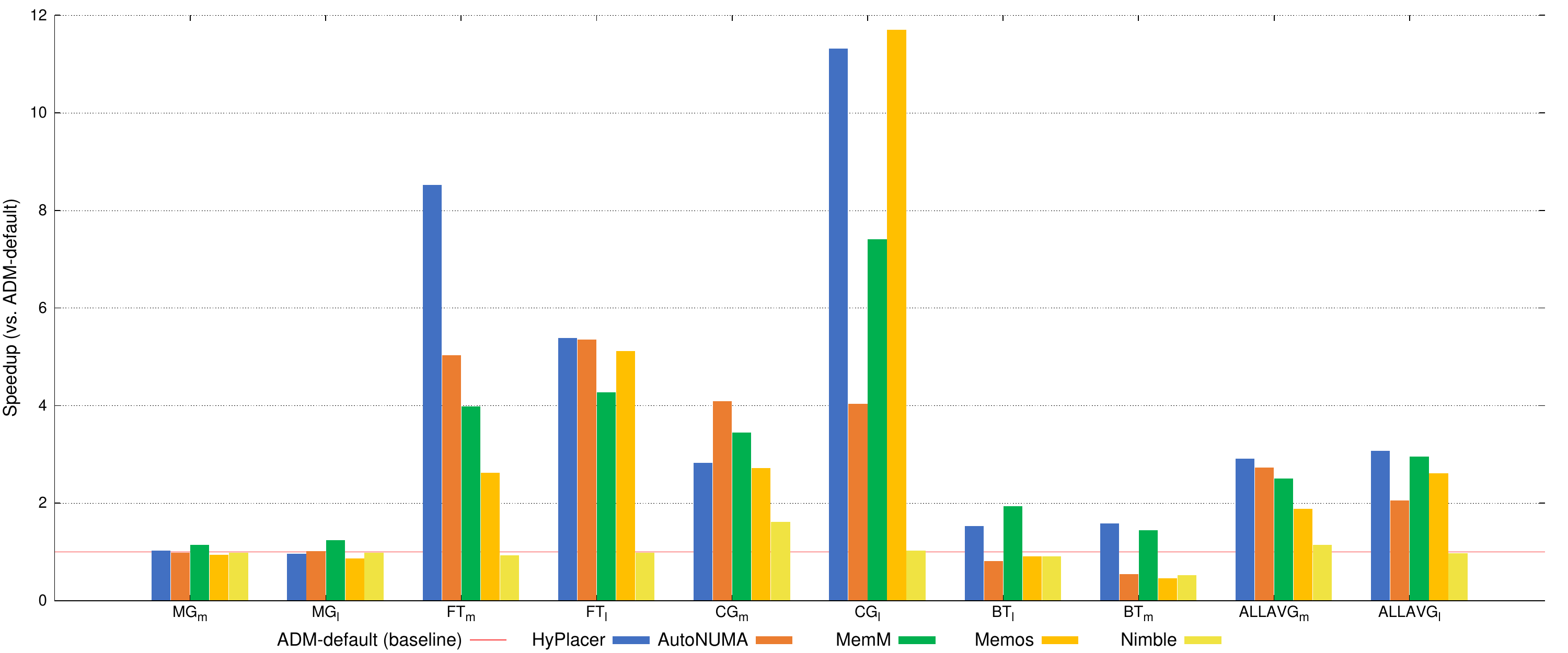}
    \caption{Throughput speedup compared to \textit{ADM-Default}.}
    \label{fig:NPB}
\end{figure*}

Each application is tested with three data set sizes: small, which totally fits
in DRAM; medium, which is approximately 1.5x DRAM size; and large, approximately 3.5x DRAM size. The small size is a worst-case scenario for which tiered page placement brings no effective advantage to a trivial static placement in DRAM.
The medium and large sizes are the most relevant ones, as they require the
supplementary memory capacity that \optane provides, hence it is in these scenarios
that tiered page placement solutions are expected to benefit the application.


For each experiment, the results we show next are averaged over 3 runs.

\begin{figure*}[t]
\centering
    \includegraphics[width=0.95\linewidth]{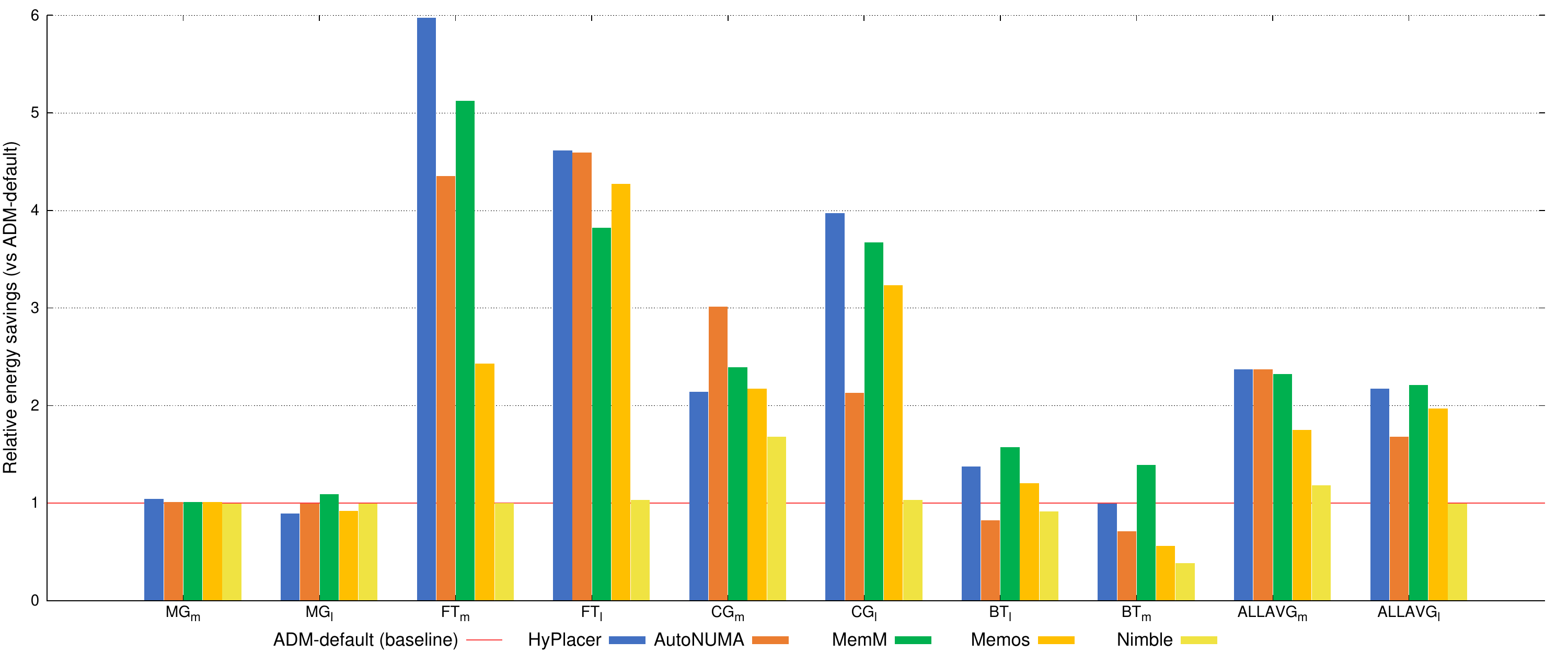}
    \caption{Per-access memory energy consumption compared to \textit{ADM-Default}.}
    \label{fig:NPB-energy}
\end{figure*}

\subsection{Performance results with large data sets}
\label{sec:bigdata}

\Cref{fig:NPB} presents a set of plots showing the performance speedup for each of the BT, FT, MG, and CG benchmarks under both medium and large size data sets compared to the \textit{ADM-default} configuration, as well as the geometric average speedup of all four benchmarks.

\rr{Maybe structure according to the systems being compared to. The next paragraph is a nice comparison to memos, but then this structure might be replicated to the other systems?}\jcm{modified the text in the next paragraph, but please complete, before it said memos was the worse, but nimble is even worse... we should also say something about MemM, I don't know what we're doing with that}

We observe that \textit{nimble} performs significantly worse than the other solutions for most of the cases, being at par or worse relative to the \textit{ADM-default} configuration. Among other factors, this can be explained by \emph{nimble}'s simplistic page selection mechanism, which exclusively considers page hotness, hence is not aware of the read/write intensity of the pages it decides to migrate across tiers. Furthermore, this also reflects a sub-optimal parameter choice, which was originally
defined based on innacurate assumptions about the real persistent memory.

The \textit{memos} solution also presents lower speedups than the remaining tested solutions in the majority of scenarios, with an average 28\% performance reduction compared to the \textit{ADM-default} configuration. This solution performs best in the very read-intensive CG workloads, but even for this case it is only better than \name for the medium-sized load. The reasons for this low performance are the poor initial memory placement policy of Memos, which allocates new pages in \optane, and also due to its ineffective bandwidth-aware promotion mechanism, which often fails to saturate DRAM throughput.

Comparing the other solutions in the medium data sets, \textit{MemM}, \textit{autonuma}, and \name  show a speedup of 2.5x, 2.3x, and 3.7x on average vs.\ \textit{ADM-default}.
The speedup increases in the large parametrizations, to 3.8x, 2.8x, and 5.4x, respectively.

We observe that \name's placement mechanisms improve throughput more than the \textit{autonuma}, despite only changing a single line of code in the kernel, and processing most of the placement decisions in a user-level process.
In all workloads, the \textit{ambix} configuration grants an average speedup of 4.6x, compared to \textit{autonuma}'s 2.6x.
Its benefit is most noticeable in the BT workloads, where \textit{autonuma} fails to improve the \textit{ADM-default} configuration, while \textit{ambix} has an average 2.25x speedup.
In the read-intensive CG workload, \textit{autonuma} grants better performance than \textit{ambix} in the medium parametrization, but falls off in the large one, with a 4x vs. 11x speedup compared to \textit{ADM-default}.


\textit{MemM} performs better than \textit{autonuma} in the majority of medium and large workloads.
However, \textit{ambix} surpasses it, having a 45\% higher average speedup in all workloads.

Besides throughput, we measured the energy consumption of each application with each
page placement approach. To measure energy consumption, we used Linux's \emph{perf} tool to sample the energy consumption of the memory system; more precisely, using \emph{perf stat} to collect measurements from event \emph{power/energy-ram} every second.

Figure~\ref{fig:NPB-energy} depicts the gains in energy for each solution for the same instances as Figure~\ref{fig:NPB}, again using the energy consumption of the \textit{ADM-default} configuration as baseline. 
Note that in Figure~\ref{fig:NPB-energy} higher is better as the values represent how many times lower the energy consumption is relative to the baseline.  

These results demonstrate that a good page placement decision can have a significant impact on the energy efficiency of a program. Furthermore, comparing these two figures, we can observe that the trends of energy gains are mostly consistent with the throughput speedup values (in Figure \ref{fig:NPB}). 

\igk{,,TO DO: some observations about the energy results confirm that energy consumption are quite correlated to throughput; point out some exceptions to that and try to justify them.`` -- It would be easier to perform this task, if we plot energy and performance on the same plot, will do it}\jcm{I agree there is one or two vlaues outside the general trends, unless you have a good insight about them, I suggest we ignore them...}

\rr{It is striking that memory mode is en par or even slightly better than Ambix on a fraction of the applications, and then quite worse than Ambix on the other subset. Why is this the case?}\jpb{There are papers that have studied the workload patterns that favor memM. I will find them and borrow their arguments/citation here.}

\begin{figure}[t]
\centering
    \includegraphics[width=0.95\linewidth]{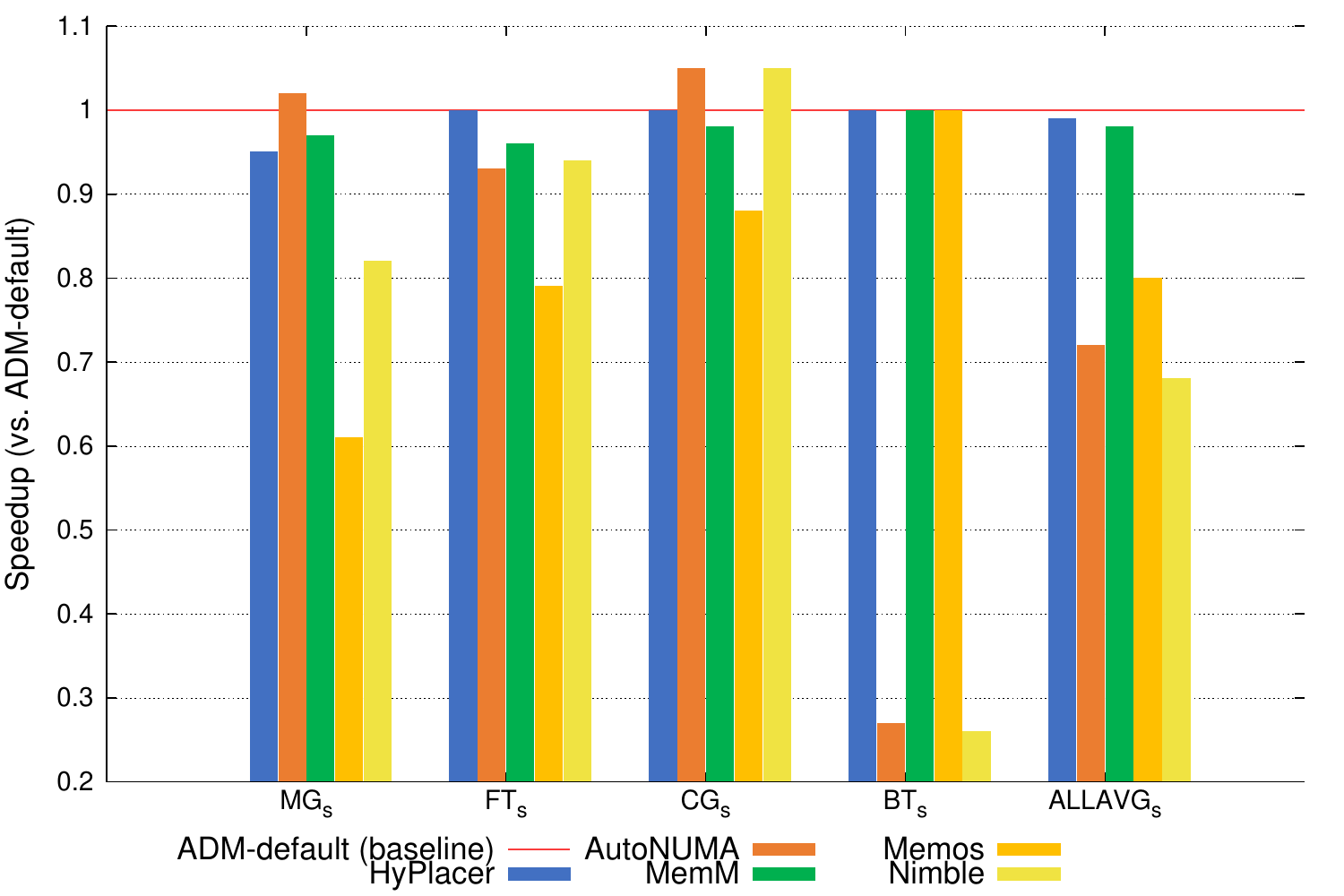}
    \caption{Throughput speedup compared to \textit{ADM-Default}, with data sets that fit entirely in DRAM.}
    \label{fig:NPB-small}
\end{figure}

\subsection{Overheads in the worst-case scenario (small data sets)}
\label{sec:eval:smalldata}

We now turn our attention to the overheads of the different solutions.
To evaluate that, we consider a worst-case scenario where the data sets of every application are small enough to fit entirely in DRAM.
Of course, in this scenario, the optimal solution consists in simply placing 
every page in DRAM, statically. For this reason, the slowdowns presented in Figure~{fig:NPB-small} over the baseline are to be expected, representing the overheads associated with each proposed solution. We observe that in most cases the results are close to the baseline. Still, \textit{ambix} presents a higher penalty for the MG and FT cases, justified by preemptive, unnecessary page migration. \jcm{please review}

\section{Conclusions}
\label{section:conclusion}

In this paper, we are the first, to our knowledge, to systematically
analyze
tiered page placement on real \optane-based systems.
Based on an initial experimental study using a real
\optane-based system, we reveal that some of the key design choices in the literature rely on important assumptions that are not verified in contemporary DRAM-\optane memory architectures.

Based on these lessons, we propose \name, a practical tool for tiered page placement in \emph{off-the-shelf} Linux-based systems equipped with DRAM+\optane.
In contrast to previous proposals, \name follows an approach guided by two main practicality principles:
1) it is tailored to the actual performance characteristics of off-the-shelf DRAM+\optane systems;
and 2) it can be seamlessly integrated into Linux with minimal kernel-mode components, while ensuring extensibility to other HMAs and other data placement policies.

We evaluate \name, as well as relevant page placement alternatives, using realistic benchmarks.
We show that \name outperforms both solutions proposed in past literature as well as placement options that are currently available in off-the-shelf \optane-equipped Linux systems, with an average speedup of 4.6x in large footprint workloads, reaching a peak improvement of 11x when compared to the default memory policy in Linux.

\jpb{maybe mention future work directions. some ideas: integration with profile-based static placement tool to give hints to ambix about the default placement of each page; combine with proposals for cross-socket (NUMA) placement; combine with shadow-copy-based general frameworks for persistent data strucutures (whose shadow pages are statically placed in DRAM but can actually be managed by a tiered page placement policy).}

\newpage

\section*{Acknowledgments}
This work was partially supported by FCT (UIDB/50021/2020), and EU's H2020 R\&I programme (EPEEC project, GA 801051).


\bibliographystyle{IEEEtranS}
\bibliography{refs}

\end{document}